\documentclass[twocolumn,showpacs,showkeys,superscriptaddress]{revtex4}

\usepackage{amsmath}
\usepackage{bbold}
\usepackage{amsfonts}
\usepackage{amssymb}
\usepackage{pbsi}
\usepackage[T1]{fontenc}
\usepackage{hyperref}
\usepackage{xcolor}
\usepackage{graphicx}
\usepackage{subfig}
\usepackage{float}
\usepackage{color}

\begin{document}

\title{London-like tensor modes of gravitational waves in cosmic string cosmology}

\author{Claudio Aravena-Plaza}
\email{c_aravena@ug.uchile.cl (corresponding author)}
\affiliation{Departamento de F\'{\i}sica, Facultad de Ciencias, Universidad de Chile,
Santiago 7800003, Chile.}

\author{V\'ictor Mu\~noz}
\affiliation{Departamento de F\'{\i}sica, Facultad de Ciencias, Universidad de Chile,
Santiago 7800003, Chile.}

\author{Felipe A. Asenjo}
\affiliation{Facultad de Ingenier\'ia y Ciencias,
Universidad Adolfo Ib\'a\~nez, Santiago 7491169, Chile.}

\date{\today}
\begin{abstract}
From a classical analysis, we show that gravitational waves in a cosmological medium with equation of state $\omega=-1/3$ can follow a London-like equation, implying that some gravitational wave solutions present a decay for certain wavelengths. This  scenario, corresponding to a cosmic string cosmology, induces an attenuation temporal scale on the gravitational wave propagation. We discuss on
how these solutions impose a limit on the wavelength of the waves that can propagate,  which depends on the type of spatial curvature  and the energy density content of this type of cosmology.
 \end{abstract}


\maketitle

\section{Introduction}

From the linearization of the Einstein equations for a Friedmann–Lema\^itre–Robertson–Walker (FLRW) cosmological background metric, the scalar, vector and tensor modes for gravitational perturbations are straightforwardly obtained \cite{weinberg}. The tensor modes are of particular importance because they describe gravitational waves on this cosmology \cite{weinberg}. The detection of these  waves is important because of the idea of an inflationary primitive Universe producing primordial gravitational waves that, if measured, would be a substantial contribution to understanding the evolution of our Universe \cite{osti_981884}.

In general, the study of tensor modes for a FLRW background metric can be performed exactly. Their dynamical evolution is determined by the equation of state of the cosmological content of the Universe, mainly described by a perfect fluid with pressure $p$ and its energy density 
$\rho$, and thus, equation of state $\omega=p/\rho$. The different values of $\omega$ characterize the evolution of the scale factor of the universe. The distinction is usually made between oscillatory solutions with a cutoff value of $\omega >-1/3$, where the solutions oscillate and damp, and with $\omega < -1/3$, where the oscillation occurs and then freezes at some value.

On the other hand, the characterization of  the Universe for $\omega=-1/3$ lies within the so-called dark energy cosmologies, and consists of lineal-like forms of energy, called the cosmic strings. A cosmic string could arise as a topological defect that occurred during a phase transition early in the Universe \cite{Kibble_1976, Bormotova, vilenkin}. Even if the strings formed before the last epoch of inflation, they were thought to have been diluted by inflation and would be too few to be cosmologically observable. However strong evidence for a common spectrum process has recently been reported by the NANOGrav Collaboration, which was interpreted as a stochastic gravitational wave background (SGWB) in the framework of cosmic strings \cite{PhysRevD.107.042003}. The importance of this measurement opens the field of observation of primordial background waves that would allow us to obtain information about the Universe at this stage \cite{nanograv}. 
Cosmic strings can produce primordial gravitational waves, such as gravitational wave bursts and SGWBs \cite{Siemens_2007, https://doi.org/10.48550/arxiv.2207.03510, https://doi.org/10.48550/arxiv.2006.13872, CMB, Emond_2022,Caprini_2018}. 
 In general, to show that a universe with a cosmic string can produce primordial gravitational waves, models described by some action are used, for example,  using the Nambu-Goto action \cite{auclair2021window}, or
Lagrangian models
\cite{Emond_2022}.
 
 Unlike those works, here we carry out a first order tensor mode analysis for gravitational waves in a Universe filled with a perfect fluid with equation of state $\omega=-1/3$. We show that dynamics follows a London equation. This implies that when the medium has this equation of state, there exist conditions for which gravitational waves cannot propagate through it, as they fully damp on a specific temporal scale. London equations have been studied for linear corrections of gravitational fields in flat spacetime \cite{Ummarino}, but not in the context of cosmological gravitational waves.


The studied gravitational wave modes in this work correspond to spacetime perturbations of  
the FLRW  background metric, which is described by the spacetime interval
\begin{eqnarray}
\label{eq:1}
ds^{2}=g_{\mu\nu}dx^{\mu}dx^{\nu}=a^2\left(-d\eta^2+\gamma_{ij}dx^{i}dx^{j}\right)\, ,
\end{eqnarray}
where $a=a(t)$ is the scale factor of the Universe, and 
$x^\mu=(t,x^i)$ (with $\mu,\nu=0,1,2,3$, and $i,j=1,2,3$). We have chosen the speed of light $c=1$. In the following we describe the dynamics in terms of the cosmological time $\eta=\int dt/a$. The three-dimensional spatial metric $\gamma_{ij}$ can be represented in spherical coordinates as  
$\gamma_{ij}dx^{i}dx^{j}=dD^{2}+D_{A}^{2}d\Omega$,
where $D$ is the comoving distance and $D_{A}=K^{-1/2}\sin(K^{1/2}D)$ is the angular diameter distance. 
$K$ is a constant representing the spatial curvature of the Universe ($K=0$ for spatially flat,   $K=1$ for closed Universe, and $K=-1$ for open Universe).

By linearizing the Einstein equations with this cosmological background metric, the equation for the tensor mode,  in  absence of corrections to the stress anisotropy, is obtained from Eq.~(\ref{eq:11}) of Appendix A.
By defining the effective gravitational wave $y$ by the relation 
$H_{T}=y/a$, then we get
\begin{equation}
\label{eq:2}
\ddot{y}+\left(k^{2}+2K -\frac{\ddot{a}}{a}\right)y=0\, ,
\end{equation}
where the overdot stands for the conformal time derivative, and 
we have assumed the spatial behavior of the wave is sinusoidal
$\nabla^{2}y=-k^{2} y$, with $k$ the wavenumber of the gravitational wave.
Also,  the background FLRW metric evolves following the equation ${\ddot{a}}/{a}=({4\pi G}/{3})a^{2}\rho-K$, whereas the fluid content satisfies the continuity equation $\dot{\rho}/\rho=-3\left(1+\omega\right)\left(\dot{a}/a\right)$, whose solution is 
$\rho =\rho_{0}a^{-3(1+\omega)}$, where $\rho_0$ is initial energy density. 
Substituting in Eq.~(\ref{eq:2}), we finally obtain 
\begin{equation}
\label{eq:3}
\ddot{y}+\left(k^{2}+3K-\frac{4\pi G}{3}a^{-1-3\omega}\rho_{0} (1-3\omega)\right)y=0\, .
\end{equation}
By knowing the type of content of the Universe, and thus, how it evolves, by means of $a$, diverse forms of gravitational propagation can be found from this equation \cite{Caldwell_1993, grishchuk1974amplification, Peter_K_S_Dunsby_1997}.

\section{London-like behavior}
\label{results}

Historically, the London equation is a phenomenological classical, way to relate the current in a superconductor with  the electric and magnetic fields inside and outside of it. The idea is that electrons move freely inside a superconductor, following equations detailed in Appendix B. Thereby, 
we obtain that the magnetic field $\mathbf{B}$ can penetrate inside a superconductor, and satisfies the London equation  $     \nabla^{2}\mathbf{B}=\mathbf{B}/{l^{2}}$, 
where $l=\sqrt{m/4\pi n e^{2}}$ is called the penetration length of the magnetic field. Here, $m$ is electron mass, and $n$ and $e$ are the electron density and charge respectively. The solution of this equation describes classically the Meissner effect, for an exponentially screened magnetic field
inside a superconductor, depending on the medium through the penetration length  \cite{london,tinkham}.

In this section we show that there exists a cosmological scenario where gravitational wave behavior resembles this exponentially screened behavior, i.e.,  Eq.~(\ref{eq:3}) acquires a London-like form.
In Eq.~(\ref{eq:3}), let us consider the cosmic string  equation of state $\omega=-1/3$. In that case, there is no explicit dependence on $a$,  and it becomes simply 
\begin{equation}
\label{eq:5}
\ddot{y}+\left(k^{2}+3K-\frac{8\pi G}{3}\rho_{0} \right)y=0\, .
\end{equation}
This equation presents a London-like behavior for gravitational waves as it can be written as 
\begin{equation}
\label{eq:6}
\ddot{y}=\frac{1}{\Omega^2}y\, ,
\end{equation}
where
\begin{equation}
\label{eq:7}
\Omega = \left(\frac{8\pi G}{3}\rho_{0}-k^{2}-3K\right)^{-1/2}\,
\end{equation}
is an attenuation conformal cosmological temporal-scale, that depends on the initial energy density content of this cosmic string universe $\rho_0$, the spatial curvature, and the wavelength of the gravitational field wave.  This attenuation temporal-scale is valid only when
the gravitational wave wavenumber satisfies
\begin{equation}
\frac{8\pi G}{3}\rho_{0}-3K > k^{2}\, .  
\label{eq:8}
\end{equation}

Instead of propagating, now the gravitational wave field is exponentially screened in certain cosmological time $\Omega$. This is seen from the physical solution of Eq.~(\ref{eq:6}), which reads
\begin{equation}
\label{eq:8b}
y(\eta)=y_{0}\exp\left(-\frac{\eta}{\Omega}\right)\, 
\end{equation}
where $y_{0}$ is an integration constant, representing the initial amplitude of the wave. This behavior for cosmological gravitational wave fields is the temporal analogue to the London behavior of superconductivity. However, the temporal attenuation of the gravitational perturbation does not only depend on the medium (as in the electromagnetic London case) but also on its own wave features. In effect, similar to the reflection of long wavelength waves in magnetized plasmas, gravitational waves with small wavenumbers can satisfy Eq.~\eqref{eq:8}, and thus it is expected that no long wavelength gravitational waves  propagate in this era; they are all exponentially screened in a conformal cosmological temporal scale of the order of $\Omega$.

\section{Discussion}
The above conditions for tensor modes, obtained for a cosmology with a Universe filled with matter with $\omega=-1/3$ equation of state, determine the passage of gravitational waves through it or not. This is the typical result derived from a London-like equation, such as Eq.~\eqref{eq:6}.
Under condition \eqref{eq:8}, the modes gets damped in conformal time. This  only occurs for gravitational waves. For example, electromagnetic waves always propagate as plane waves in isotropic cosmology,  suffering no damping from the medium.

As a proper estimation, let us consider the spatially flat Universe case. Then, any gravitational wave perturbation with wavelength $\lambda=2\pi/k$ larger than $\sqrt{3\pi/(2 G \rho_0)}$, becomes time screened from propagation. The energy density of cosmic strings at formation can be put in the form $\rho_0\sim \mu/L_0^2$, where $\mu$ is the linear mass density of the string and $L_0$ is a characteristic length scale of their pseudo-Brownian trajectories \cite{vilenkin}. Thereby, any temporal attenuated gravitational wave in this cosmology will have a wavelength that satisfies
\begin{equation}
    \frac{\lambda}{L_0}>\sqrt{\frac{3\pi}{2G\mu}}\, .
\end{equation}
If the linear mass density of cosmic strings lie between
 $\mu\sim 10^{-4}$ g/cm on the electroweak scale, to $\mu\sim 10^{22}$ g/cm for grand unification strings  \cite{vilenkin}, then, we have that $\lambda/L_0> 2.5\times 10^{16}$--$10^{3}$, from  the electroweak scale to grand unification strings.
On the contrary, when $k^{2}$ does not satisfy  condition \eqref{eq:8}, then cosmlogical gravitational waves can travel through this medium. 
\\
Recently, it has been argued that gravitational waves originating from  cosmic cues of the SGWB type have been detected \cite{PhysRevLett.126.041304, BUCHMULLER2020135914}. Although this interpretation is controversial \cite{PhysRevD.103.103512}, this goes in the direction of advancing the measurement of an analogue of the microwave background, but with primordial gravitational waves \cite{auclair2021window,PhysRevD.103.043023}. 
Our result imposes wavelength conditions that establish which gravitational waves are damped, and therefore detectable or undetectable. An estimate of the value of the density $\rho_{0}$ can even be obtained by knowing the specific wavelength constraints and the respective attenuation time scale for this specific type of Universe.

\begin{acknowledgments}
CAP thanks to ANID No.~7322/2020. This project has been financially supported by FONDECyT under
contracts No.~1201967 (VM), and  No.~1230094 (FAA). 
\end{acknowledgments}

\section{APPENDIX}
\subsection{Tensorial gravitational modes in FLRW cosmology}
The gravitational tensor modes $H_{T}$ for a spacetime perturbation of the FLRW metric, without  dissipative corrections to the inertia tensor, follow the wave equation \citep{weinberg}  
\begin{equation}
\label{eq:11}
\nabla^{2}H_{T}-{a^{2}}\left(\dfrac{d^{2}}{dt^{2}}H_{T}\right)-3{a}\left(\dfrac{d}{dt}a\right)\left(\dfrac{d}{dt}H_{T}\right)=0\, .
\end{equation}
The evolution of these modes as a function of the cosmological time $\eta$ can be obtained by noticing that
$\dot{a}\equiv{da}/{d\eta}= a\, {da}/{dt}$. Incorporating curvature constant $K$ of the Universe, we obtain
\begin{equation}
\label{eq:12}   
-\nabla^{2}H_{T} + \ddot{H}_{T} + 2\frac{\dot{a}}{a}\dot{H}_{T} + 2K H_{T}=0,
\end{equation}
where $\dot{H}_{T}\equiv{d}H_{T}/d\eta$.

\subsection{London equation}

The London equation 
was formulated to have a classical phenomenological explanation of superconductivity. It is
derived from the equations
\begin{equation}
\label{eq:17}
    \nabla \times \mathbf{j}_{s}=-\frac{n_{s}e^{2}}{m}\mathbf{B}\, ,\qquad \nabla\times {\mathbf B}=4\pi {\mathbf j}_s\, ,
\end{equation}
where $\mathbf{j}_{s}$ is the current in a superconductor, $\mathbf{B}$ is the   magnetic field,             $e$ is the electron charge, $m$ is the electron mass, and $n_{s}$ is the number density of particles moving within the superconductor.
By combining them, we obtain  the London equation.

\bibliographystyle{unsrt}
\bibliography{london_waves}

\begin{thebibliography}{10}

\bibitem{weinberg}
S.~Weinberg.
\newblock {\em "Cosmology"}.
\newblock Oxford, (2008).

\bibitem{osti_981884}
L.~Krauss, S.~Dodelson, and S.~Meyer.
\newblock Primordial gravitational waves and cosmology.
\newblock {\em Science}, 328, (2010).

\bibitem{Kibble_1976}
T.~W.~B. Kibble.
\newblock Topology of cosmic domains and strings.
\newblock {\em Journal of Physics A: Mathematical and General}, 9(8), (1976).

\bibitem{Bormotova}
I.~Bormotova and E.~Kopteva.
\newblock Friedmann cosmological models with various equations of state of
  matter.
\newblock {\em Ukrainian Journal of Physics}, 61(9), (2016).

\bibitem{vilenkin}
A.~Vilenkin.
\newblock Cosmic strings and domain walls.
\newblock {\em Physics Reports}, 121(5), (1985).

\bibitem{PhysRevD.107.042003}
Y.~M. Wu, Z.~C. Chen, and Q.~G. Huang.
\newblock Search for stochastic gravitational-wave background from massive
  gravity in the nanograv 12.5-year dataset.
\newblock {\em Phys. Rev. D}, 107, Feb (2023).

\bibitem{nanograv}
M.~Benetti, G.~Leila, and S.~Vagnozzi.
\newblock Primordial gravitational waves from nanograv: A broken power-law
  approach.
\newblock {\em Physical review D: Particles and fields}, 105, (2021).

\bibitem{Siemens_2007}
X.~Siemens, V.~Mandic, and J.~Creighton.
\newblock Gravitational-wave stochastic background from cosmic strings.
\newblock {\em Physical review letters}, 98(4), (2007).

\bibitem{https://doi.org/10.48550/arxiv.2207.03510}
P.~Auclair, S.~Blasi, V.~Brdar, and K.~Schmitz.
\newblock Gravitational waves from current-carrying cosmic strings, (2022).

\bibitem{https://doi.org/10.48550/arxiv.2006.13872}
R.~Zhou and L.~Bian.
\newblock Gravitational waves from cosmic strings and first-order phase
  transition, (2020).

\bibitem{CMB}
A.~Bodas and R.~Sundrum.
\newblock Large primordial fluctuations in gravitational waves from phase
  transitions, (2022).

\bibitem{Emond_2022}
W.~T. Emond, S.~Ramazanov, and R.~Samanta.
\newblock Gravitational waves from melting cosmic strings.
\newblock {\em Journal of Cosmology and Astroparticle Physics}, (2022)(01),
  (2022).

\bibitem{Caprini_2018}
C.~Caprini and D.~G. Figueroa.
\newblock Cosmological backgrounds of gravitational waves.
\newblock {\em Classical and Quantum Gravity}, 35(16), (2018).

\bibitem{auclair2021window}
P.~Auclair, K.~Leyde, and D.~A. Steer.
\newblock A window for cosmic strings.
\newblock {\em Journal of Cosmology and Astroparticle Physics}, (2023)(04),
  2023.

\bibitem{Ummarino}
G.~A. Ummarino and A.~Gallerati.
\newblock Superconductor in a weak static gravitational field.
\newblock {\em Eur. Phys. J. C}, 77(549), (2017).

\bibitem{Caldwell_1993}
R.~R. Caldwell.
\newblock Green{\textquotesingle}s functions for gravitational waves in {FRW}
  spacetimes.
\newblock {\em Physical Review D}, 48(10), (1993).

\bibitem{grishchuk1974amplification}
L.~Grishchuk.
\newblock Amplification of gravitational waves in an istropic universe.
\newblock {\em Zh. Eksp. Teor. Fiz}, 67, (1974).

\bibitem{Peter_K_S_Dunsby_1997}
P.~Dunsby, B.~Bassett, and G.~Ellis.
\newblock Covariant analysis of gravitational waves in a cosmological context.
\newblock {\em Classical and Quantum Gravity}, 14(5), (1997).

\bibitem{london}
H.~London F.~London.
\newblock The electromagnetics equations of the supraconductor.
\newblock {\em Proc. R. Soc. Lond.}, 149(866), (1935).

\bibitem{tinkham}
M.~Tinkham.
\newblock {\em Introduction to Superconductivity}.
\newblock Dover Publications, Inc., (1996).

\bibitem{PhysRevLett.126.041304}
J.~Ellis and M.~Lewicki.
\newblock Cosmic string interpretation of nanograv pulsar timing data.
\newblock {\em Phys. Rev. Lett.}, 126, (2021).

\bibitem{BUCHMULLER2020135914}
W.~Buchmuller, V.~Domcke, and Schmitz K.
\newblock From nanograv to ligo with metastable cosmic strings.
\newblock {\em Physics Letters B}, 811, (2020).

\bibitem{PhysRevD.103.103512}
J.~Blanco-Pilladoand~K. Olum and J.~Wachter.
\newblock Comparison of cosmic string and superstring models to nanograv
  12.5-year results.
\newblock {\em Phys. Rev. D}, 103, (2021).

\bibitem{PhysRevD.103.043023}
K.~Martinovic, P.~Meyers, M.~Sakellariadou, and N.~Christensen.
\newblock Simultaneous estimation of astrophysical and cosmological stochastic
  gravitational-wave backgrounds with terrestrial detectors.
\newblock {\em Phys. Rev. D}, 103, (2021).

\end{thebibliography}

\end{document}